\documentclass[12pt]{iopart}
\usepackage{graphicx,bm,url}

\newcommand{\EQ}{\begin{equation}}
\newcommand{\EN}{\end{equation}}
\newcommand{\EQA}{\begin{eqnarray}}
\newcommand{\ENA}{\end{eqnarray}}
\newcommand{\eq}[1]{(\ref{#1})}
\newcommand{\EEq}[1]{Equation~(\ref{#1})}
\newcommand{\Eq}[1]{Equation~(\ref{#1})}
\newcommand{\Eqs}[2]{Equations~(\ref{#1}) and~(\ref{#2})}

\newcommand{\Sec}[1]{Section~\ref{#1}}

\newcommand{\Fig}[1]{Figure~\ref{#1}}
\newcommand{\FFig}[1]{Figure~\ref{#1}}

\newcommand{\bra}[1]{\langle #1\rangle}

{}
{}
{}

{}
{}
\newcommand{\meanEMF}{\overline{\mbox{\boldmath ${\cal E}$}}{}}{}
{}
{}
{}
{}
\newcommand{\meanAA}{\overline{\mbox{\boldmath $A$}}{}}{}
\newcommand{\meanBB}{\overline{\mbox{\boldmath $B$}}{}}{}
{}
\newcommand{\meanFF}{\overline{\mbox{\boldmath $F$}}{}}{}
{}
{}
{}
{}
{}
\newcommand{\meanJJ}{\overline{\mbox{\boldmath $J$}}{}}{}
\newcommand{\meanUU}{\overline{\bm{U}}}
{}
{}
{}

\newcommand{\meanU}{\overline{U}}

\newcommand{\meanJ}{\overline{J}}

\newcommand{\hatAA}{\hat{\bm{A}}}
\newcommand{\hatBB}{\hat{\bm{B}}}
{}

{}
{}

\newcommand{\alphaK}{\alpha_{\it K}}
\newcommand{\alphaM}{\alpha_{\it M}}
\newcommand{\kk}{\bm{k}}
\newcommand{\pp}{\bm{p}}
\newcommand{\qq}{\bm{q}}
\newcommand{\xx}{\bm{x}}

\newcommand{\uu}{\mbox{\boldmath $u$} {}}
\newcommand{\UU}{\mbox{\boldmath $U$} {}}

\newcommand{\bb}{\mbox{\boldmath $b$} {}}
\newcommand{\BB}{\mbox{\boldmath $B$} {}}

\newcommand{\jj}{\mbox{\boldmath $j$} {}}
\newcommand{\JJ}{\mbox{\boldmath $J$} {}}
\newcommand{\SSS}{\mbox{\boldmath $S$} {}}
\newcommand{\AAA}{\mbox{\boldmath $A$} {}}

\newcommand{\ee}{\mbox{\boldmath $e$} {}}

\newcommand{\FF}{\mbox{\boldmath $F$} {}}

\newcommand{\nab}{\mbox{\boldmath $\nabla$} {}}

\newcommand{\oo}{\mbox{\boldmath $\omega$} {}}

\newcommand{\EMF}{\mbox{\boldmath ${\cal E}$} {}}

\newcommand{\ii}{{\rm i}}

\newcommand{\dd}{{\rm d} {}}

\def\ga{\mathrel{\mathchoice {\vcenter{\offinterlineskip\halign{\hfil
$\displaystyle##$\hfil\cr>\cr\sim\cr}}}
{\vcenter{\offinterlineskip\halign{\hfil$\textstyle##$\hfil\cr>\cr\sim\cr}}}
{\vcenter{\offinterlineskip\halign{\hfil$\scriptstyle##$\hfil\cr>\cr\sim\cr}}}
{\vcenter{\offinterlineskip\halign{\hfil$\scriptscriptstyle##$\hfil\cr>\cr\sim\cr}}}}}
\def\Rm{R_{\rm m}}

\def\cs{c_{\rm s}}

\def\kf{k_{\rm f}}

\def\urms{u_{\rm rms}}

\def\etat{\eta_{\rm t}}
\def\etatz{\eta_{\rm t0}}

\def\etaT{\eta_{\rm T}}

\def\Beq{B_{\rm eq}}

\def\half{{\textstyle{1\over2}}}

\def\onethird{{\textstyle{1\over3}}}

\newcommand{\G}{\,{\rm G}}

\newcommand{\Mm}{\,{\rm Mm}}
\newcommand{\Mx}{\,{\rm Mx}}

\newcommand{\yjgr}[3]{ #1, {J.\ Geophys.\ Res.,} {#2}, #3}

\newcommand{\yapj}[3]{ #1, {ApJ,} {#2}, #3}

\newcommand{\yapjl}[3]{ #1, {ApJ,} {#2}, #3}

\newcommand{\yan}[3]{ #1, {Astron.\ Nachr.,} {#2}, #3}

\newcommand{\yana}[3]{ #1, {A\&A,} {#2}, #3}

\newcommand{\ygafd}[3]{ #1, {Geophys.\ Astrophys.\ Fluid Dyn.,} {#2}, #3}

\newcommand{\yjfm}[3]{ #1, {J.\ Fluid Mech.,} {#2}, #3}

\newcommand{\ypf}[3]{ #1, {Phys.\ Fluids,} {#2}, #3}

\newcommand{\ypp}[3]{ #1, {Phys.\ Plasmas,} {#2}, #3}

\newcommand{\yjetp}[3]{ #1, {Sov.\ Phys.\ JETP,} {#2}, #3}

\newcommand{\yprs}[3]{ #1, {Proc.\ Roy.\ Soc.\ Lond.,} {#2}, #3}

\newcommand{\yprl}[3]{ #1, {Phys.\ Rev.\ Lett.,} {#2}, #3}

\newcommand{\ymn}[3]{ #1, {MNRAS,} {#2}, #3}

\newcommand{\ypre}[3]{ #1, {Phys.\ Rev.\ E,} {#2}, #3}

\newcommand{\yjour}[4]{ #1, {#2}, {#3}, #4}

\newcommand{\ybook}[3]{ #1, {#2} (#3)}

\begin{document}
\title[Magnetic helicity in astrophysical dynamos]
{The critical role of magnetic helicity in astrophysical large-scale dynamos}
\author{Axel Brandenburg}
\address{Nordita, Roslagstullsbacken 23, 10691 Stockholm, Sweden; and\\
Department of Astronomy, Stockholm University, 10691 Stockholm, Sweden
}

\begin{abstract}
The role of magnetic helicity in astrophysical large-scale dynamos is
reviewed and compared with cases where there is no energy supply and an
initial magnetic field can only decay.
In both cases magnetic energy tends to get redistributed to larger scales.
Depending on the efficiency of magnetic helicity fluxes, the decay of
a helical field can speed up.
Likewise, the saturation of a helical dynamo can speed up through magnetic
helicity fluxes.
The astrophysical importance of these processes is reviewed in the context
of the solar dynamo and an estimated upper limit for the magnetic helicity
flux of $10^{46}\Mx^2/\mbox{cycle}$ is given.
\end{abstract}

\section{Introduction}

Self-excited dynamo action refers to the instability of a plasma in
a non-magnetic equilibrium state to amplify magnetic fields within the
framework of resistive magnetohydrodynamics (MHD).
Charge separation effects are assumed absent, i.e.\ no battery-type
effects are explicitly involved, although they do play a role in producing
a weak initial seed magnetic field that is needed to provide a perturbation
to the otherwise field-free initial state.
A dynamo instability may occur when the magnetic Reynolds number is large
enough, i.e.\ the fluid motions and the scale of the domain are large enough.
This instability is normally a linear one, but some dynamos
are subcritical and require then a finite-amplitude initial field.
In the linear case one speaks about slow and fast dynamos depending on
whether or not the growth rate of the dynamo scales with resistivity.
For fast dynamos the growth rate scales with the rms velocity of the
flow, which is turbulent in most cases.
Dynamos saturate when the magnetic energy becomes comparable with the
kinetic energy.
Some of the kinetic energy is then channelled through the magnetic
energy reservoir and is eventually dissipated via Joule heating.

There are several applications where one considers non self-excited
dynamo action.
Examples can be found in magnetospheric physics and in plasma physics
where one is interested in the electromotive force induced by a flow
passing through a given magnetic field.
Later in this paper we will discuss the reversed field pinch (RFP)
because of its connection with the $\alpha$ effect that plays
an important role in astrophysical dynamos.
The physics of the RFP has been reviewed by Ortolani \& Schnack (1993)
and, in a broader context with reference to astrophysical large-scale
dynamos, by Ji \& Prager (2002) and Blackman \& Ji (2006).

The $\alpha$ effect is one of a few known mechanisms able to generate
large-scale magnetic fields, i.e.\ fields whose typical length scale
is larger than the scale of the energy carrying motions.
It was also one of the first discussed mechanisms able to produce
self-excited dynamo action at all.
Indeed, Parker (1955) showed that the swirl of a convecting flow under
the influence of the Coriolis force can be responsible for producing
a systematically oriented poloidal magnetic field from a toroidal field.
The toroidal field in turn is produced by the shear from the differential
rotation acting on the poloidal field.
Parker's paper was before Herzenberg (1958) produced the first existence
proof of dynamos.
Until that time there was a serious worry that Cowling's (1933) anti-dynamo
theorem might carry over from two-dimensional fields to three-dimensional
fields, as is evident from sentimental remarks made by Larmor (1934).

Larmor (1919) proposed the idea of dynamo action in the
astrophysical context nearly 100 years ago.
Nowadays, with the help of computers, it is quite easy to solve the
induction equation in three dimensions in simple geometries and obtain
self-excited dynamo solutions with as little as $16^3$ mesh points
using, for example, the sample ``\texttt{kin-dynamo}'' that comes with
the {\sc Pencil Code} (\url{http://pencil-code.googlecode.com}).

In addition to dynamos in helical flows, which can generate large-scale
fields, there are also dynamos in non-helical flows that produce only
small-scale fields.
This possibility was first addressed by Batchelor (1950) based on
the analogy between the induction equation and the vorticity equation.
Again, this was not yet very convincing at the time.
The now accepted theory for small-scale dynamos was first proposed by
Kazantsev (1968) and the first simulations were produced by
Meneguzzi et al.\ (1981).
Such simulations are computationally somewhat more demanding and require
at least $64^3$ mesh points (or collocation points in spectral schemes).
In the past few years this work has intensified (Cho et al.\ 2002,
Schekochihin et al.\ 2002, Haugen 2003).
We will not discuss these dynamos in the rest of this paper.
Instead, we will focus on large-scale dynamos.
More specifically, we focus here on a special class of large-scale
dynamos, namely those where kinetic helicity plays a decisive role
(the so-called $\alpha$ effect dynamos).
Nevertheless, we mention at this point two other mechanisms that
could produce large-scale magnetic field without net helicity.
One is the incoherent $\alpha$--shear effect that was originally proposed
by Vishniac \& Brandenburg (1997) to explain the occurrence of large-scale
magnetic fields in accretion discs, and later also for other astrophysical
applications (e.g., Proctor 2007).
It requires the presence of shear, because otherwise only small-scale
magnetic fields would be generated (Kraichnan 1976, Moffatt 1978).
The other mechanism is the shear--current effect of
Rogachevskii \& Kleeorin (2003), which can operate if the turbulent
magnetic diffusion tensor is anisotropic, so the mean electromotive
force from the turbulence is given by $-\eta_{ij}\meanJ_j$ such that the
sign of $\eta_{ij}\meanU_{i,j}$ is positive, and that this quantity is
big enough to overcome resistive effects.
Here, a comma denotes partial differentiation, $\meanUU$ is the mean flow,
$\meanJJ$ is the mean current density,
and summation over repeated indices is assumed.
This effect too requires shear, because otherwise $\eta_{ij}\meanU_{i,j}$
would be zero.
Simulations show large-scale dynamo action in the presence
of just turbulence and shear, and without net helicity, but there
are indications that this process may also be just the result of incoherent
$\alpha$--shear dynamo action (Brandenburg et al.\ 2008a).

Throughout this paper, overbars denote suitable spatial averages
over one or two coordinate directions.
Furthermore, we always assume that the scale of the energy-carrying eddies
is at least three times smaller than the scale of the domain.
We refer to this property as ``scale separation''.
In this sense, scale separation is a natural requirement, because we
want to explain the occurrence of fields on scales large compared with
the scale of the energy-carrying motions.
Scale separation has therefore nothing to do with a gap in the kinetic
energy spectrum, as is sometimes suggested.

Much of the work on $\alpha$ effect dynamos has been done in the
framework of analytic approximations.
However, this is only a technical aspect that is unimportant for the
actual occurrence of large-scale fields under suitable conditions.
This has been demonstrated by numerical simulations, as will be discussed
below.

\section{Helical large-scale dynamos}

A possible way of motivating the physics behind helical dynamo action
is the relation to the concept of an inverse turbulent cascade.
This particular idea was first proposed by Frisch et al.\ (1975) and is
based on the conservation of magnetic helicity,
\EQ
H_{\rm M}=\int_V\AAA\cdot\BB\,\dd V,
\EN
where $\AAA$ is the magnetic vector potential and $\BB=\nab\times\AAA$
is the magnetic field in a volume $V$.

\begin{figure}[t!]\begin{center}
\includegraphics[width=.6\columnwidth]{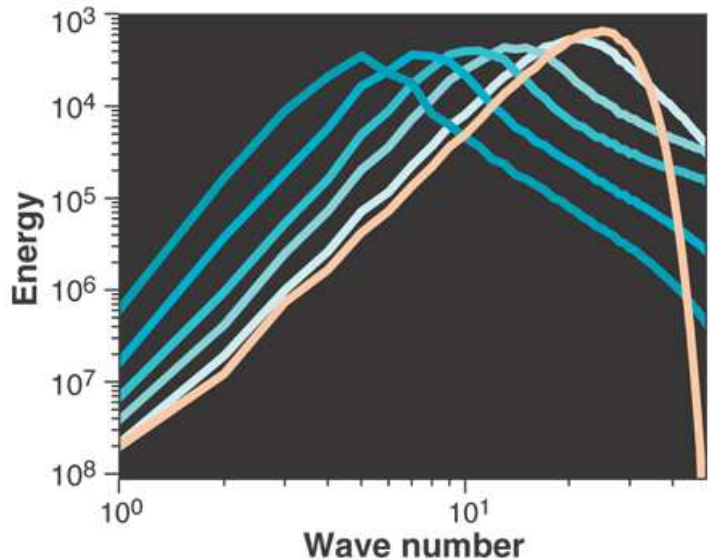}
\end{center}\caption[]{
Magnetic energy spectra at different times
(increasing roughly by a factor of 2).
The curve with the right-most location of the peak corresponds to
the initial time, while the other lines refer to later times (increasing
from right to left). Note the
propagation of spectral energy to successively smaller wavenumbers $k$,
i.e.\ to successively larger scales.
Adapted from Christensson et al.\ (2001).
}\label{InvCasc}\end{figure}

It is convenient to define spectra of magnetic energy and magnetic
helicity, $E_{\rm M}(k)$ and $H_{\rm M}(k)$, respectively.
As usual, these spectra are obtained by calculating the three-dimensional
Fourier transforms of magnetic vector potential and magnetic field,
$\hatAA_{\kk}$ and $\hatBB_{\kk}$, respectively,
and integrating $|\hatBB_{\kk}|^2$ and
the real part of $\hatAA_{\kk}\cdot\hatBB_{\kk}^*$ over shells of constant
$k=|\kk|$ to obtained $E_{\rm M}(k)$ and $H_{\rm M}(k)$, respectively.
(Here, an asterisk denotes complex conjugation.)
These spectra are normalized such that
$\int E_{\rm M}(k)\,\dd k=\bra{\BB^2}/2\mu_0$ and
$\int H_{\rm M}(k)\,\dd k=\bra{\AAA\cdot\BB}$ for $k$ from 0 to $\infty$,
where angular brackets denote volume averages over a periodic domain
and $\mu_0$ is the vacuum permeability.
Using the Schwartz inequality one can then derive the so-called realizability
condition,
\EQ
k|H_{\rm M}(k)|/2\mu_0\leq E_{\rm M}(k).
\EN
For fully helical magnetic fields with (say) positive helicity,
i.e.\ $H_{\rm M}=2\mu_0 E_{\rm M}(k)/k$, one can show that energy
and magnetic helicity cannot cascade directly, i.e.\ the interaction
of modes with wavenumbers $\pp$ and $\qq$ can only produce fields whose
wavevector $\kk=\pp+\qq$ has a length that is equal or smaller than the maximum
of either $|\pp|$ or $|\qq|$ (Frisch et al.\ 1975), i.e.\
\EQ
|\kk|\leq\max(|\pp|,|\qq|).
\label{kpq}
\EN
This means that magnetic helicity and magnetic energy are transformed to
progressively larger length scales.
A clear illustration of this can be seen in decaying helical turbulence.
\FFig{InvCasc} shows magnetic energy spectra from a simulation of
Christensson et al.\ (2001) at different times for a
case where the initial magnetic field was fully helical and had a spectrum
proportional to $k^4$ with a resolution cutoff near the largest possible
wavenumber.
Note that the entire spectrum appears to shift to the left, i.e.\ toward
larger length scales, in an approximately self-similar fashion.
The details of the argument that led to \Eq{kpq} are due to
Frisch et al.\ (1975), and can also be found in the reviews by
Brandenburg et al.\ (2002) and Brandenburg \& Subramanian (2005a).

\begin{figure}[t!]
\centering\includegraphics[width=.5\textwidth]{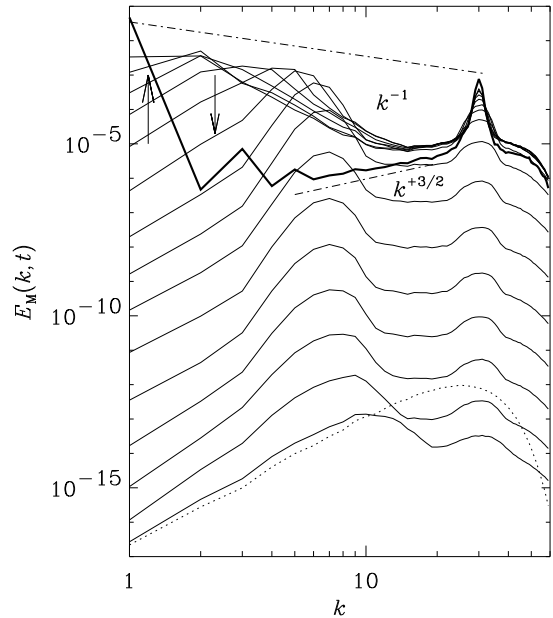}\caption{
Magnetic energy spectra for a run with forcing at $k=30$. The times,
in units of $(\cs k_1)^{-1}$,
range from 0 (dotted line) to 10, 30, ..., 290 (solid lines). The thick
solid line gives the final state at $\cs k_1 t=1000$,
corresponding to $\urms\kf t\approx2000$ turnover times.
Here, $\cs$ is the sound speed and $\urms$ is the turbulent rms velocity.
Note that at early times
the spectra peak at $k_{\max}\approx7k_1$. The $k^{-1}$ and $k^{+3/2}$
slopes are given for orientation as dash-dotted lines.
Adapted from Brandenburg (2001).
}\label{FpMkt2}\end{figure}

In the non-decaying case, when the flow is driven by energy input at some
forcing wavenumber $\kf$, the inverse cascade is clearly seen if there is
sufficient scale separation, i.e.\ if $\kf$ is large compared with the
smallest wavenumber $k_1$ that fits into a domain of size $L=2\pi/k_1$.
An example is shown in \Fig{FpMkt2}, where kinetic energy is injected
at the wavenumber $\kf=30k_1$.
It is evident that there are two local maxima of spectral magnetic
energy, one at the forcing wavenumber $\kf$, and another one at a smaller
wavenumber that we call $k_{\rm m}$, which is near $7k_1$ in \Fig{FpMkt2}.
During the kinematic stage the entire spectrum moves upward, with the
spectral energy increasing at the same rate at all wavenumbers.
Eventually, when the field has reached a certain level, the spectrum
begins to change its shape and the second local maximum at $k_{\rm m}$
moves toward smaller values, suggestive of an inverse cascade.
However, a more detailed analysis (Brandenburg 2001) shows that the
energy transfer is nonlocal, i.e.\ most of the energy is transferred
directly from the forcing wavenumber to the wavenumber where most of
the mean field resides.
This suggests that we have merely a nonlocal inverse {\it transfer} rather than
a proper inverse {\it cascade}, where the energy transfer would be local in
spectral space.

The position of the local maximum can readily be explained by mean-field
dynamo theory with an $\alpha$ effect.
The evolution equation of such a dynamo is
\EQ
{\partial\meanBB\over\partial t}=\nab\times\alpha\meanBB+\etaT\nabla^2\meanBB,
\label{MFD}
\EN
where $\alpha$ is a pseudo-scalar, $\etaT=\eta+\etat$ is the sum of
microscopic Spitzer resistivity and turbulent resistivity\footnote{Note
that resistivity and magnetic diffusivity differ by a $\mu_0$ factor.
Here, we always mean the magnetic diffusivity, although we use the two
names sometimes interchangeably.}, and an overbar denotes a suitably
defined spatial average (e.g.\ planar average).
Assuming $\meanBB=\hatBB_{\kk}\exp(\lambda t+\ii\kk\cdot\xx)$ with
eigenfunction $\hatBB_{\kk}$, one finds the dispersion relation to be
(Moffatt 1978)
\EQ
\lambda(k)=|\alpha|k-\etaT k^2,
\EN
where $k=|\kk|$.
The maximum growth rate is attained for a value of $k$ where
$\dd\lambda/\dd k=0$, i.e.\ for $k=k_{\rm m}=\alpha/2\etaT$.
The migration of the spectral maximum to smaller $k$ can then be
explained as the result of a suppression of $\alpha$.
In other words, as the dynamo saturates, $\alpha$ decreases, and so does
$k_{\rm m}$, which corresponds to the spectral maximum moving to the left.

It should be noted that there is also the possibility of a suppression
of $\etat$, which would work the other way, so that this interpretation
might not work.
Indeed, there are arguments for a suppression of $\etat$ that would be as
strong as that of $\alpha$, but this applies only to the two-dimensional
case and has to do with the conservation of the mean-squared vector
potential in that case (Gruzinov \& Diamond 1994).
Simulations also find evidence that in three dimensions the suppression
of $\alpha$ is stronger than that of $\etat$ (Brandenburg et al.\ 2008b).

In the following we discuss in detail the role played by magnetic helicity.
This has been reviewed extensively in the last few years
(Ji \& Prager 2002, Brandenburg \& Subramanian 2005, Blackman \& Ji 2006).

\section{Slow saturation}

In a closed or periodic domain, the saturation of a helical large-scale
dynamo is found to be
resistively slow and the final field strength is reached with a time
behavior of the form (Brandenburg 2001)
\EQ
\bra{\meanBB^2}(t)=\Beq^2{k_{\rm m}\over\kf}
\left[1-e^{-2\eta k_{\rm m}^2(t-t_{\rm s})}\right]
\quad\mbox{for $t>t_{\rm s}$},
\label{SlowSat}
\EN
where $\Beq$ is the equipartition field strength and $t_{\rm s}$ is
the time when the slow saturation phase begins.
We emphasize that it is the microscopic $\eta$ that enters \Eq{SlowSat},
and that the relevant length scale, $2\pi/k_{\rm m}$, is that of the
large-scale field, so the saturation behavior is truly very slow.

The reason for this slow saturation behavior is related to the conservation
of magnetic helicity, which obeys the evolution equation
(e.g., Ji et al.\ 1995, Ji 1999)
\EQ
{\dd\over\dd t}\bra{\AAA\cdot\BB}=-2\eta\mu_0\bra{\JJ\cdot\BB}
-\bra{\nab\cdot\FF_{\rm H}}.
\label{dABdt}
\EN
where $\FF_{\rm H}$ is the magnetic helicity flux, but for the periodic
domain under consideration we have $\nab\cdot\FF_{\rm H}=0$.
Clearly, in the final state we have then $\bra{\JJ\cdot\BB}=0$.
This can only be satisfied for nontrivial helical fields if
small-scale and large-scale fields have values of opposite sign,
but equal magnitude, i.e.\ $\bra{\jj\cdot\bb}=-\bra{\meanJJ\cdot\meanBB}$,
where $\BB=\meanBB+\bb$ and $\JJ=\meanJJ+\jj$ are the decompositions of
magnetic field and current density into mean and fluctuating parts.
Here we choose to define mean fields as one- or two-dimensional coordinate
averages.
Examples include planar averages such as $xy$, $yz$, or $xz$ averages
in a periodic Cartesian domain, as well as one-dimensional averages
such as $y$ or $\phi$ averages in Cartesian or spherical domains,
$(r,\theta,\phi)$, where the other two directions are non-periodic.

\EEq{SlowSat} can be derived under the assumption that large-scale and
small-scale fields are fully helical with
$\meanJJ\cdot\meanBB=k_{\rm m}^2\meanAA\cdot\meanBB=\mp k_{\rm m}\meanBB^2$
and $\bra{\jj\cdot\bb}=\pm\kf\bra{\bb^2}\approx\pm\kf\Beq^2$,
where upper and lower signs refer to positive and negative
helicity of the small-scale turbulence and we have assumed that
the small-scale field has already reached saturation, i.e.\
$\bra{\bb^2}\approx\Beq^2\equiv\mu_0\bra{\rho\uu^2}$, where $\rho$ is
the density and $\uu=\UU-\meanUU$ is the fluctuating velocity.

A mean-field theory that obeys magnetic helicity conservation was
originally developed by Kleeorin \& Ruzmaikin (1982) and has recently
been applied to explaining slow saturation (Field \& Blackman 2002,
Blackman \& Brandenburg 2002, Subramanian 2002).
The main idea is that the $\alpha$ effect has two contributions
(Pouquet et al.\ 1976),
\EQ
\alpha=\alphaK+\alphaM,
\label{alphaKM}
\EN
where $\alphaK=-\onethird\tau\overline{\oo\cdot\uu}$ is the usual kinetic
$\alpha$ effect related to the kinetic helicity,
with $\oo=\nab\times\uu$ being the vorticity, and
$\alphaM=\onethird\tau\overline{\jj\cdot\bb}/\rho_0$ is a magnetic
$\alpha$ effect that can, for example, be produced by the growing
magnetic field in an attempt to conserve magnetic helicity.
(Here, $\rho_0$ is an average density, but we note that there is at present
no adequate theory for compressible systems with nonuniform density.)

Note that $\alphaM$ is related to the small-scale current helicity
and hence to the small-scale magnetic helicity which, in turn, obeys
an evolution equation similar to \Eq{dABdt}, but with an additional
production term, $2\meanEMF\cdot\meanBB$, that arises from mean-field
theory via $\meanEMF=\alpha\meanBB-\etat\mu_0\meanJJ$, and thus from
the $\alpha$ effect itself.
The equation for $\alphaM$ has then the form
\EQ
{\partial\alphaM\over\partial t}=-2\etat\kf^2
\left({\meanEMF\cdot\meanBB\over\Beq^2}+{\alphaM\over\Rm}\right)
-\nab\cdot\meanFF_\alpha,
\label{dalphaMdt}
\EN
where $\Rm=\etat/\eta$ is a measure of the ratio of turbulent to
microscopic magnetic diffusivity.
With $\etat=\urms/3\kf$ (Sur et al.\ 2008) we can relate this to
the more usual definition for the magnetic Reynolds number,
$\tilde\Rm=\urms/\eta\kf$, via $\tilde\Rm=3\Rm$.
Furthermore, we have allowed for the possibility of fluxes of magnetic and
current helicities that also lead to a flux of $\alphaM$.
Such fluxes are primarily important in inhomogeneous domains and especially
in open domains where one can have an outward helicity flux (Ji 1999).

The use of \Eq{alphaKM} is sometimes criticized because it is based on
a closure assumption.
Indeed, there are questions regarding the meaning of the term
$\overline{\jj\cdot\bb}$ and whether it really applies to the actual
field, or the field in the unquenched case.
This has been discussed in detail in a critical paper by
R\"adler \& Rheinhardt (2007).
Part of this ambiguity can already be clarified in the low conductivity limit.
Sur et al.\ (2007) have shown that one can express $\alpha$ either
completely in terms of the helical properties of the velocity field or,
alternatively, as the sum of two terms, a so-called kinetic $\alpha$
effect and an oppositely signed term proportional to the helical part
of the small scale magnetic field.
However, it is fair to say that the problem is not yet completely understood.
The strongest argument in favor of \Eqs{alphaKM}{dalphaMdt} is that
they reproduce catastrophic (i.e.\ $\Rm$-independent) quenching of $\alpha$
and that this approach has led to the prediction that such quenching can be
alleviated by magnetic helicity fluxes.
This prediction has subsequently been tested successfully on various
occasions (Brandenburg 2005, K\"apyl\"a et al.\ 2008).
Finally, it should be noted that \Eq{alphaKM} has also been confirmed
directly using turbulence simulations (Brandenburg et al.\ 2005c, 2007).

The idea to model dynamo saturation and suppression of $\alpha$ by solving
a dynamical equation for $\alphaM$ is called dynamical quenching.
In addition to the resistively slow saturation behavior described by
\Eq{SlowSat}, this approach has also been applied to decaying turbulence
with helicity (Yousef et al.\ 2003; Blackman \& Field 2004), where the
conservation of magnetic helicity results in a slow-down of the decay.
This can be modeled by an $\alpha$ effect that offsets the turbulent
decay proportional to $\etat k^2$ such that the decay rate becomes nearly
equal to the resistive value, $\eta k^2$.
This is explained in detail in the following section.

\section{Decay in a Cartesian domain}
\label{Sdynquench}

In the context of driven turbulence, the properties of solutions of a
decaying helical magnetic field were studied earlier by
Yousef et al.\ (2003), who found that for fields with
$\meanBB^2/B_{\rm eq}^2\ga R_{\rm m}^{-1}$ the decay of $\meanBB$ is slowed
down and can quantitatively be described by the dynamical quenching model.
This model applies even to the case where the turbulence is nonhelical and
where there is initially no $\alpha$ effect in the usual sense.
However, the magnetic contribution to $\alpha$ is still non-vanishing,
because the $\alphaM$ term is driven by the helicity of the
large-scale field.

To demonstrate this quantitatively, Yousef et al.\ (2003) have adopted
a one-mode approximation with $\meanBB=\hatBB(t)\exp(\ii k_1z)$,
and used the mean-field induction equation together with the dynamical
$\alpha$-quenching formula \eq{dalphaMdt},
\EQ
{\dd\hatBB\over\dd t}=\ii\kk_1\times\hat{\EMF}
-\eta k_1^2\hatBB,
\label{ODE1}
\EN
\EQ
{\dd\alpha\over\dd t}=-2\etat k_{\rm f}^2
\left({{\rm Re}(\hat{\EMF}^*\cdot\hatBB) \over B_{\rm eq}^2}
+{\alpha\over\Rm}\right),
\label{ODE2}
\EN
where the flux term is neglected,
$\hat{\EMF}=\alpha\hatBB-\eta_{\rm t}\ii\kk_1\times\hatBB$
is the electromotive force, and $\kk_1=(0,0,k_1)$.

\begin{figure}[t!]
\centering\includegraphics[width=0.5\textwidth]{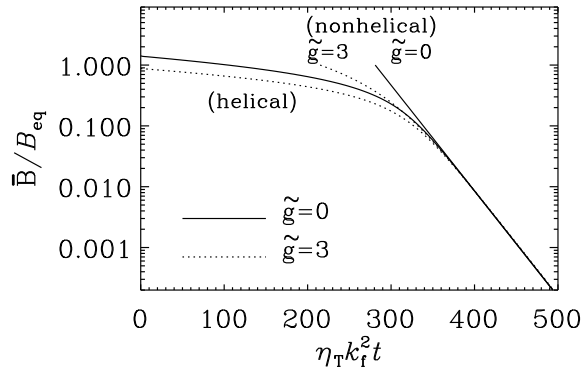}\caption{
Dynamical quenching model with helical and nonhelical initial fields,
and an additional $\etat$ quenching function,
$\etat=\etatz/(1+\tilde{g}|\meanBB|/\Beq)$.
The quenching parameters are $\tilde{g}=0$ (solid line) and 3 (dotted line).
The graph for the nonhelical cases has been shifted in $t$ so that one
sees that the decay rates are asymptotically equal at late times.
The value of $\eta_{\rm T}$ used to normalize the abscissa is based
on the unquenched value.
Adapted from Yousef et al.\ (2003).
}\label{Fpselected_decay}\end{figure}

\FFig{Fpselected_decay} compares the evolution of $\meanBB/B_{\rm eq}$
for helical and nonhelical initial conditions, $\hatBB\propto(1,\ii,0)$
and $\hatBB\propto(1,0,0)$, respectively.
In the case of a nonhelical field, the decay rate is not quenched at all,
but in the helical case quenching sets in for
$\meanBB^2/B_{\rm eq}^2\ga R_{\rm m}^{-1}$.
The onset of quenching at $\meanBB^2/B_{\rm eq}^2\approx R_{\rm m}^{-1}$
is well reproduced by the simulation.
In the nonhelical case, however, some weaker form of quenching sets in
when $\meanBB^2/B_{\rm eq}^2\approx1$.
We refer to this as standard quenching (e.g.\ Kitchatinov et al.\ 1994)
which is known to be always present; see \Eq{SlowSat}.
Blackman \& Brandenburg (2002) found that, for a range of different
values of $R_{\rm m}$, $\tilde{g}=3$ results in a good description of
the simulations of cyclic $\alpha\Omega$-type dynamos that were reported
by Brandenburg et al.\ (2002.).

\section{Relevance to the reversed field pinch}

The dynamical quenching approach has been applied to modeling
the dynamics of the reversed field pinch, where one has an initially
helical magnetic field of the form
\EQ
\meanBB=\hatBB\pmatrix{0\cr J_1(kr)\cr J_0(kr)},
\EN
where we have adopted cylindrical coordinates $(r,\theta,z)$.
In a cylinder of radius $R$ such an initial field becomes kink unstable
when $kR\ga\pi$.
Both laboratory measurements (e.g.\ Caramana \& Baker 1984) and numerical
simulations (Ho et al.\ 1989) confirm the idea that the field-aligned
current leads to kink instability, and hence to small-scale turbulence
and thereby to the emergence of $\etat$ and $\alpha$.

\begin{figure}[t]\begin{center}
\includegraphics[width=.6\columnwidth]{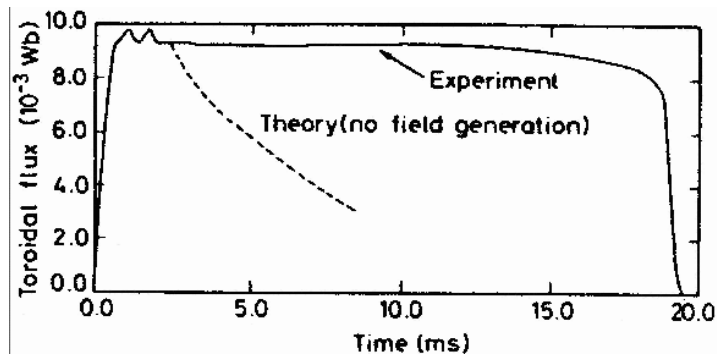}
\end{center}\caption[]{
Time evolution of toroidal flux in the RFP experiment of
Caramana \& Baker (1984) compared with a calculation with no dynamo effect.
The decay around $t=18$ms is due to termination of the applied electric field.
Courtesy of E. J. Caramana and D. A. Baker.
}\label{toroidal_flux}\end{figure}

Just as in the case discussed in \Sec{Sdynquench}, the emergence of $\alpha$
slows down the decay in such a way that the toroidal field remains nearly
constant and is maintained against resistive decay; see \Fig{toroidal_flux}.
The details of this mechanism have been discussed by Ji \& Prager (2002).
In particular, they show that the parallel electric field cannot be
balanced by the resistive term alone, and that there must be an additional
component resulting from small-scale correlations of velocity and magnetic
field, $\overline{\uu\times\bb}$, that explain the observed profiles of
mean electric field and mean current density.
Furthermore, the parallel electric field reverses sign near the edge
of the device, while the parallel current density does not.
Again, this can only be explained by additional contributions from
small-scale correlations of velocity and magnetic field.
The RFP experiment also shows that magnetic helicity evolves on time scales
faster than the resistive scale, which is only compatible with the presence
of a finite magnetic helicity flux divergence (Ji et al.\ 1995, Ji 1999).

\section{Magnetic helicity in realistic dynamos}

Astrophysical dynamos saturate and evolve on dynamical time scales and
are thus not resistively slow.
Current research shows that this can be achieved by expelling
magnetic helicity from the domain through helicity fluxes.
This is why we have allowed for the $\nab\cdot\meanFF_\alpha$ term
in \Eq{dalphaMdt}.
Since the magnetic $\alpha$ effect is proportional to the current
helicity of the fluctuating field, the $\meanFF_\alpha$ flux should
be proportional to the current helicity flux of the
fluctuating field.

The presence of the flux term generally lowers the value of
$|\alphaM|$, and since the $\alphaM$ term quenches the total
value of $\alpha$ ($=\alphaK+\alphaM$), the effect of this helicity flux
is to alleviate an otherwise catastrophic quenching.
Indeed, in an open domain and without a flux divergence, the $\alphaM/\Rm$
term in \Eq{dalphaMdt} can result in ``catastrophically low'' saturation field
strengths that are by a factor $\Rm^{1/2}$ smaller than the equipartition
field strength (Gruzinov \& Diamond 1994; Brandenburg \& Subramanian 2005b).

Magnetic helicity obeys a conservation law and is therefore
conceptually easier to tackle than current helicity.
However, there is the difficulty of gauge dependence of magnetic
helicity density and its flux.
It is therefore safer to work with the current helicity, which is
also the quantity that enters in \Eqs{alphaKM}{dalphaMdt}.
However, more work needs to be done to establish the connection between
the two approaches.

Over the past 10 years there has been mounting evidence that the Sun
sheds magnetic helicity (and hence current helicity) through coronal
mass ejections and other events.
Understanding the functional form of such fluxes is very much a matter
of ongoing research (Subramanian \& Brandenburg 2004, 2006,
Brandenburg et al.\ 2009).
In the following section we present a simple calculation that allows
us to estimate the amount of magnetic helicity losses required for the
solar dynamo to work.

\section{Estimating the required magnetic helicity losses}

In order to estimate the magnetic helicity losses required to alleviate
catastrophic quenching we make use of the relation between the $\alpha$ effect
and the divergence of the current helicity flux
(Brandenburg \& Subramanian 2005a),
\EQ
\alpha={\alphaK+\Rm\left[
\eta_{\rm t}\mu_0\meanJJ\cdot\meanBB/B_{\rm eq}^2
-\nab\cdot\meanFF_{\rm C}/(2 k_{\rm f}^2B_{\rm eq}^2)
-\dot\alpha/(2\eta_{\rm t} k_{\rm f}^2)\right]
\over1+\Rm\meanBB^2/B_{\rm eq}^2},
\label{QuenchExtra2}
\EN
where $\dot\alpha=\partial\alpha/\partial t$ and $\meanFF_{\rm C}$
is the mean flux of current helicity from the small-scale field,
$\overline{(\nab\times\ee)\times(\nab\times\bb)}$, where $\ee$ is
the fluctuating component of the electric field; see also
Subramanian \& Brandenburg (2004).
In the steady-state limit and at large $\Rm$ we have
\EQ
\alpha\approx
\eta_{\rm t}{\mu_0\meanJJ\cdot\meanBB\over\meanBB^2}
-{\nab\cdot\meanFF_{\rm C}\over2 k_{\rm f}^2\meanBB^2}.
\label{QuenchExtra3}
\EN
We neglect the $\meanJJ\cdot\meanBB$ term, because the catastrophic
quenching in dynamos with boundaries has never been seen to be alleviated
by this term (Brandenburg \& Subramanian 2005b).
Thus, we use \Eq{QuenchExtra3} to estimate $\nab\cdot\meanFF_{\rm C}$ as
\EQ
\nab\cdot\meanFF_{\rm C}=2\alpha k_{\rm f}^2\meanBB^2.
\EN
Next, we take the volume integral over one hemisphere, i.e.\
\EQ
{\cal L}_{\rm C}\equiv\oint_{2\pi}\meanFF_{\rm C}\cdot\dd\SSS
=\int\nab\cdot\meanFF_{\rm C}\,\dd V
={2\pi\over3}R^3\bra{2\alpha k_{\rm f}^2\meanBB^2},
\EN
where ${\cal L}_{\rm C}$ is the ``luminosity'' or ``power'' of current helicity.
We estimate $\bra{\alpha}$ using $\alpha\Omega$ dynamo theory
which predicts that (e.g., Robinson \& Durney 1982, see also
Brandenburg \& Subramanian 2005a)
\EQ
\alpha k_1\Delta\Omega\approx\omega_{\rm cyc}^2,
\EN
where $\omega_{\rm cyc}=2\pi/T_{\rm cyc}$ is the cycle frequency
of the dynamo, $T_{\rm cyc}$ is the 22 year cycle period of the Sun,
$\alpha$ is assumed constant over each hemisphere, and $\Delta\Omega$
is the total latitudinal shear, i.e.\ about $0.3\Omega$ for the Sun.
There is obviously an uncertainty in relating local values of
$\alpha$ to volume averages.
However, if we do set the two equal, we obtain at least an upper limit
for ${\cal L}_{\rm C}$.
We may then relate this to the luminosity of {\it magnetic} helicity,
${\cal L}_{\rm H}$, that we assume to be proportional to ${\cal L}_{\rm C}$
via
\EQ
{\cal L}_{\rm C}=\kf^2{\cal L}_{\rm H}.
\label{LCLH}
\EN
With this we find for the total magnetic helicity loss over half a
cycle (one 11 year cycle)
\EQ
\half{\cal L}_{\rm H}T_{\rm cyc}\leq{4\pi\over3}R^3
{\omega_{\rm cyc}^2\over k_1\Delta\Omega}T_{\rm cyc}\bra{\meanBB^2}
={4\pi\over3}LR^3{\omega_{\rm cyc}\over\Delta\Omega}\bra{\meanBB^2},
\EN
where we have estimated $k_1=2\pi/L$ for the relevant wavenumber of the
dynamo in terms of the thickness of the convection zone $L$.
Inserting now values relevant for the Sun, $L=200\Mm$, $R=700\Mm$,
$\omega_{\rm cyc}/\Delta\Omega=10^{-2}$, and $\meanBB\sim300\G$,
we obtain
\EQ
\half{\cal L}_{\rm H}T_{\rm cyc}\leq10^{46}\Mx^2/\mbox{cycle}.
\label{estimate}
\EN
This is comparable to earlier estimates based partly on observations
(Berger \& Ruzmaikin 2000; DeVore 2000) and partly on turbulence simulations
(Brandenburg \& Sandin 2004), but we recall that \Eq{estimate} is only
an upper limit.
Furthermore, the connection between current helicity and magnetic helicity
assumed in relation \eq{LCLH} is quite rough and has only been seriously
confirmed under isotropic conditions.
In addition, it is not clear that the gauge-invariant magnetic helicity
flux defined by Berger \& Field (1984) and Finn \& Antonsen (1985)
is actually the quantity of interest.
Following earlier work of Subramanian \& Brandenburg (2006), the
gauge-invariant magnetic helicity is not in any obvious way related
to the current helicity which, in turn, is related to the density
of the flux linking number and is at least approximately equal to
the magnetic helicity in the Coulomb gauge.

In summary, although magnetic helicity is conceptually advantageous
in that it obeys a conservation equation, the difficulty in dealing
with a gauge-dependent quantity can be quite serious.
Moreover, as emphasized before, it is really the current helicity that
is primarily of interest, so it would be useful to shift attention from
magnetic helicity fluxes to current helicity fluxes.

\section{Conclusions}

In this review we have attempted to highlight the importance of
magnetic helicity in modern nonlinear dynamo theory.
Much of the early work on the reversed field pinch since the
mid 1980s now proves to be extremely relevant in view of the
possibility of resistively slow saturation by a self-inflicted
build-up of small-scale current helicity, $\overline{\jj\cdot\bb}$,
as the dynamo produces large-scale current helicity, $\meanJJ\cdot\meanBB$.
Since this concept is not yet universally accepted, the additional evidence
from the reversed field pinch experiment can be quite useful.

The interpretation of resistively slow saturation has led to the
proposed solution that magnetic helicity fluxes (or current helicity
fluxes) are responsible for removing excess small-scale magnetic
helicity from the system.
This allows the dynamo to reach saturation levels that can otherwise
be $\Rm^{1/2}$ times smaller than the equipartition value given
by the kinetic energy density of the turbulent motions.
The consequences of this prediction have been tested in direct simulations
(Fig.~3 of Brandenburg 2005 and Fig.~17 of K\"apyl\"a et al.\ 2008),
confirming thereby ultimately basic aspects of incorporating the
magnetic helicity equation into mean-field models.
However, more progress is needed in addressing questions regarding the
relative importance of magnetic and current helicity fluxes through the
surface compared to diffusive fluxes across the equator
(Brandenburg et al.\ 2009).

We reiterate that the reversed field pinch experiment is not
{\it directly} relevant to the self-excited dynamo, but rather
its nonlinear saturation mechanism.
So far, successful self-excited dynamo experiments have only been
performed with liquid sodium (Gailitis et al.\ 2000;
Stieglitz \& M\"uller 2001; Monchaux et al.\ 2007).
This may change in future given that one can usually achieve much
higher magnetic Reynolds numbers in plasmas than in liquid metals
(Spence et al.\ 2009).
It might then, for the first time, be possible to address experimentally
questions regarding the relative importance of magnetic and current
helicity fluxes for dynamos.

\subsection*{Acknowledgments}

I thank the referees for detailed and constructive comments
that have helped to improve the presentation.
This work was supported in part by the Swedish Research Council,
grant 621-2007-4064, and the European Research Council under the
AstroDyn Research Project 227952.

\end{document}